\documentclass[aps,prl,nopacs,twocolumn]{revtex4}

\usepackage{amsmath,amssymb,graphicx,aas_macros}
\usepackage[T1]{fontenc}
\usepackage[utf8]{inputenc}
\usepackage{graphicx}
\usepackage{color}

\usepackage[unicode=true,pdfusetitle,
 bookmarks=true,bookmarksnumbered=false,bookmarksopen=false,
 breaklinks=false,pdfborder={0 0 0},backref=false,colorlinks=true]
 {hyperref}
\hypersetup{
 citecolor=blue,filecolor=blue,linkcolor=blue,urlcolor=blue}

\newcommand{\be}{\begin{equation}}
\newcommand{\ee}{\end{equation}}
\newcommand{\bea}{\begin{eqnarray}}
\newcommand{\eea}{\end{eqnarray}}
\newcommand{\bean}{\begin{eqnarray*}}
\newcommand{\eean}{\end{eqnarray*}}

\newcommand{\bk}{{\mathbf k}}

\newcommand{\RR}{{\cal R}}
\newcommand{\HH}{{\cal H}}

\newcommand{\OO}{{\cal O}}

\newcommand{\de}{\delta}
\newcommand{\De}{\Delta}
\newcommand{\ep}{\epsilon}

\newcommand{\La}{\Lambda}

\newcommand{\Om}{\Omega}

\newcommand{\ra}{\rightarrow}

\def\id{{\rm 1\kern -2.5pt I}} 

\newcommand{\vx}{{\mathbf x}}
\newcommand{\ct}{{\tau}} 

\begin{document}

\title{Does small scale structure significantly affect cosmological dynamics?}
\author{Julian Adamek$^1$}
\email{julian.adamek@unige.ch}
\author{Chris Clarkson$^2$}
\author{Ruth Durrer$^1$}
\author{Martin Kunz$^{1,3}$}
\affiliation{1-- D\'epartement de Physique Th\'eorique \& Center for Astroparticle Physics, Universit\'e de Gen\`eve, Quai E.\ Ansermet 24, 1211 Gen\`eve 4, Switzerland.\\
2-- Astrophysics, Cosmology and Gravity Centre \& Department of Mathematics and Applied Mathematics, University of Cape Town, Rondebosch 7701, South Africa.\\
3-- African Institute for Mathematical Sciences, 6 Melrose Road, Muizenberg, 7945, South Africa.}

\begin{abstract} 

The large-scale homogeneity and isotropy of the universe is generally thought to imply a well defined background cosmological model.
It may not. Smoothing over structure adds in an extra contribution, transferring power from small scales up to large. 
Second-order perturbation theory implies that the effect is small, but suggests that formally the perturbation series may not converge.
The amplitude of the effect is actually determined by the ratio of the Hubble scales at matter-radiation equality and today~-- which
are entirely unrelated. This implies that a universe with significantly lower temperature today \emph{could} have significant backreaction
from more power on small scales, and so provides the ideal testing ground for understanding backreaction. We investigate this using two different N-body numerical simulations~-- a 3D Newtonian
and a 1D simulation which includes all relevant relativistic effects. We show that while perturbation theory predicts an increasing backreaction as more initial small-scale power is added,
in fact the virialisation of structure saturates the backreaction effect at the same level independently of the equality scale. This implies that backreaction is a small
effect independently of initial conditions. Nevertheless, it may still contribute at the percent level to certain
cosmological observables and therefore it cannot be neglected in precision cosmology.
 
\end{abstract}


\maketitle

\paragraph{Introduction}

Our understanding of cosmological structure formation at late times comes mainly from Newton's theory of gravity.
This ignores effects which must appear when using General Relativity. 
The effects come in a variety of forms, from dynamical effects such as frame-dragging which alter the metric at the percent
level~\cite{Lu:2008ju,Green:2010qy,Chisari:2011iq,Clarkson:2011zq,Green:2011wc,Adamek:2013wja,Adamek:2014xba,Bruni:2013mua}, to corrections to lensing and distances
which can be several percent~\cite{Bonvin:2005ps,Bonvin:2006en,BenDayan:2012ct,Umeh:2012pn,BenDayan:2013gc,Umeh:2014ana,Clarkson:2014pda}.
As future large surveys will reach this level of precision, it is important to determine accurately any relativistic contributions to structure formation.

A more speculative effect arises from averaging over small scale structure to reveal the large-scale dynamics of the universe. A
macroscopic theory of gravity involves backreaction terms which depend on the variance of the connection which, in principle, can be large. This
has led to speculations that backreaction could even mimic dark energy~\cite{Buchert:2011sx,Clarkson:2011zq}. Although somewhat fanciful, it highlights the importance of
understanding backreaction for an accurate interpretation of the background cosmological model. 

Averaging comes itself in different ways: 
Observations are smoothed over~-- the distance redshift relation is typically the monopole of a
much more complicated expression~\cite{BenDayan:2012ct,Clarkson:2014pda,Mustapha:1997xb,Bonvin:2005ps,Bull:2012aa,DiDio:2011gf,2012PhRvD..86b3510D}. 
Averaging Einstein's field equations gives apparent modifications to the expansion and acceleration rate for average
observers, and a modified curvature~\cite{Buchert:2002ij,Buchert:1999pq,Buchert:2007ik,Rasanen:2010wz,Rasanen:2011ki,Buchert:2011sx}. The
connection of these with observables is, however, not evident. 

\paragraph{Perturbation Theory}

The importance of the averaging problem may be estimated from perturbation theory {(see  \cite{Wetterich:2001kr} for an early investigation).}
At linear order in the standard model there is no backreaction from averaging owing to the assumed homogeneity of the initial conditions
{$\langle \Phi\rangle=0$}. Here {$\Phi$} is the Bardeen potential, with power spectrum $4\pi k^3\langle\Phi(\bk)\Phi^*(\bk')\rangle = (2\pi)^3P_{\Phi}(k)\de(\bk-\bk')$. 
A crude approximation is
\be\label{eq:Ppsi}
P_{\Phi}(k) =  \frac{9\De_{\RR}^2}{25(1+(k/k_{\rm eq})^4)} \, \mbox{ with } \De_{\RR}^2=2.2\times 10^{-9}\, ,
\ee
where $k_{\rm eq} =a_{\rm eq}H_{\rm eq}=H_0\sqrt{{2}\Om_m (1+z_{\rm eq})}$ is the comoving Hubble scale at matter-radiation equality and
$\Delta_\mathcal{R}$ is the amplitude of the dimensionless curvature perturbation~\cite{Ade:2013zuv}.   A much better approximation to the linear power spectrum which we use as initial condition for our simulation
is given in~\cite{Eisenstein:1997jh,Eisenstein:1997ik}.

At second order non-trivial corrections to the background appear. In the Hubble expansion rate, $H$, these are of order
$\langle \Phi\nabla^2\Phi\rangle\sim \langle \Phi\delta\rangle\sim \langle v^2\rangle$~\cite{Clarkson:2011zq,2005ApJ...628L...1S,2005PhRvD..71b3524K,2006NJPh....8..322K,2006MPLA...21.2997N,2007PhRvD..76h3011L,2008JCAP...01..013B,2008PhRvD..78h3531L,2009PhRvD..80h3525C,2011JCAP...03..029U,2011CQGra..28p4010C}. They give typical corrections of size
\be
\left(\frac{\De H}{H}\right)_0 \sim \left(\frac{k_{\rm eq}}{H_0}\right)^2\Delta_\mathcal{R}^2 \,,
\ee
Using $1+z_{\rm eq} =\rho_m(t_0)/\rho_{\rm rad}(t_0)  \approx 2.4\times 10^4\Theta_{2.7}^{-4}\Omega_m h^2$, $\Theta_{2.7}=T_0/2.7K$, we have (using $\Theta_{2.7}\simeq 1$ and trading off the large numerical factor in $z_{\rm eq}$ against one $\Delta_\mathcal{R}$)
\be\label{order-2}
\frac{\De H}{H}  \sim \Omega_m^2 h^2 \Delta_\mathcal{R} \sim 10^{-5}\,,
\ee
which is roughly the amplitude of first-order perturbations~-- it is a remarkable coincidence that in our universe
$\Delta_\mathcal{R} \sim (H_0/k_{\rm eq})^2$~\cite{2011CQGra..28p4010C}. If $k_\text{eq}$ were two orders of magnitude larger could one still conclude that backreaction is small? \footnote{Perturbations of the expansion rate depend on the choice of observers, and the degree of backreaction depends on this~\cite{2011JCAP...03..029U}. The experimentally measured Hubble rate is not simply related
to the expansion rate  $3H=\theta=n^\mu_{;\mu}$, where $n^\mu$ denotes the normal to the equal time hypersurface. Experiments rather measure
distance redshift relations, e.g. $d_L(z)$, and infer a Hubble rate from them via the background relation $d_L(z)=z/H_0 +\OO(z^2)$ for small
redshifts~\cite{2011CQGra..28p4010C}. Here we neglect this interesting subtlety and simply study whether changes in the expansion rate $\theta$ can become large due to clustering. At small redshifts, $z\ll 1$, this distinction is irrelevant.}

This does not necessarily settle the issue as we have to study what happens at higher orders. At third-order there are no new contributions on
average for Gaussian initial conditions. At fourth order corrections of the form~\cite{2011CQGra..28p4010C}
$\langle \nabla_i\Phi\nabla^i\Phi(\nabla^2\Phi)^2\rangle\sim \langle v^2\delta^2\rangle$ appear. Naively, this gives a correction
\be\label{order-4}
\frac{\De H}{H} \sim \left(\frac{k_{\rm eq}}{H_0}\right)^2\Delta_\mathcal{R}^2 \langle \delta^2\rangle\sim \Omega_m^2 h^2 \Delta_\mathcal{R} \langle \delta^2\rangle \,.
\ee
Estimating 
$\langle \delta^2\rangle$ depends sensitively on the modeling of small-scale modes (smoothing scale or UV cut-off), as it is divergent. Given
that it is certainly larger than $\mathcal{O}(1)$, the fourth-order contribution is larger than the second-order contribution, implying a
breakdown of perturbation theory for estimating backreaction. Even a model which is smoothed on 10\,Mpc scales has
$\langle \delta^2\rangle\sim 1$ which implies that to use perturbation theory to estimate backreaction we would be summing an infinite series with
terms all about the same amplitude. Consequently \eqref{order-2} cannot be trusted to give a good approximation to the full answer, and non-linear approaches such as
numerical integration of the full Einstein equations must be considered. 

Perturbation theory tells us that there are two scales relevant for establishing the amplitude of backreaction: the equality scale and a
smoothing scale in the UV applied to the perturbation $\Phi$. The first is a physical scale depending on the initial conditions in the early universe. The amplitude of backreaction
depends on the Hubble rate at matter-radiation equality because only after matter-radiation equality density and velocity perturbations start growing. Hence the farther in the past equality lies, the more perturbations have grown until today.
The second is a scale which must be
imposed by hand as a limitation of the model, and is present also in simulations due to their finite resolution. 

\paragraph{Numerical study}

What happens in a model where equality takes place much earlier, and more modes are available to increase the amplitude of  backreaction? How does
it depend on the smoothing scale? We conduct a numerical study which
provides a testing ground for understanding
backreaction when perturbation theory fails. We investigate the sensitivity of backreaction to the equality scale. By considering a
model with a much lower temperature today we move the onset of any backreaction effect to earlier times. More precisely,
we obtain an earlier onset of nonlinear evolution when the first modes reach $\delta\sim1$.
Tuning the numerical resolution we can also study the sensitivity 
to the smoothing scale.

Recently, some of us have found~\cite{Adamek:2013wja}, using a modified N-body code including the most important relativistic modifications, that
backreaction is indeed small and that second-order perturbation theory gives a good approximation.  Here we want to investigate whether this
remains true if we change the equality scale. 
We use a 1D numerical code which contains the key features of full general relativity and allows us to thoroughly investigate the UV dependence.
We also calculate the relevant terms with a 3D Newtonian simulation using a post-Newtonian technique. 
We shall show that the conclusion from perturbation theory is not valid and that
the effect from clustering stabilizes once
non-linearities become relevant roughly on the level of the second-order prediction.

The general relativistic 1D simulation is set up in the weak field regime. For scalar metric perturbations in longitudinal gauge, given by
$
ds^2 = a^2(\ct) \left[ -(1+2\Psi) d\ct^2 + (1-2\Phi) d\vx^2 \right] \, , \label{eq:1}
$ 
this is defined as follows:
we assume that the metric perturbations, $\Phi \sim  \Psi \sim {\cal O}(\ep) \ll 1$,
$v\sim \nabla\Phi \sim \nabla \Psi \sim {\cal O}(\ep^{1/2})$ but $\de\rho/\rho$ and $\nabla_i\nabla_j\Phi$ can be arbitrarily large. We include all terms up to order $\ep$.
This formalism is not adequate to describe black holes, but it is good on small scales as long as gravity is quasi-Newtonian. The scheme is first
order accurate on horizon scales and larger but at least second order accurate on small scales. It fully contains Newtonian gravity. We argue
that in a cosmological context it is accurate up to about $10^{-5}$ on all scales. More details about this formalism and the resulting
equations can be found in \cite{Adamek:2013wja,Adamek:2014xba}. For the purpose of this paper, the important point is that we have an improved
treatment of small scale corrections. We keep terms like $(\nabla \Phi)^2$ and $\Phi\nabla_i\nabla_j\Phi$ which can be enhanced for short modes, but we still drop terms like $\Phi^2$ which
remain small on all scales.

In linear perturbation theory the spatial average of both $\Phi$ and $\Psi$
vanishes. Including non-linear terms this is no longer the case. A homogeneous mode in $\Psi$ can always be absorbed in a redefinition of the time
coordinate, $\tau$.  This is a gauge freedom remaining within longitudinal gauge.  However, if we fix the
Friedmann equations to the zeroth-order background we cannot absorb a homogeneous $\Phi$-mode into the scale factor \footnote{We can still define our coordinates
such that the homogeneous $\Phi$-mode is zero at some particular instance in time, and we will use this freedom in order to set initial conditions.}. This would modify its evolution,
hence appear like an additional contribution to the energy momentum tensor.  Such a time dependent homogeneous mode, denoted $\Phi_0(t)$, leads to a
modification of the Hubble parameter, $\HH \ra \HH -\Phi_0' =n^\mu_{;\mu}/3$, where $\HH$ denotes the comoving Hubble parameter and ${}'=d/d\tau$.
Within our approximation scheme,  $\Phi_0$ obeys 
\be
3\HH\Phi'_0 - \frac{5}{2} \langle\Phi \nabla^2 \Phi\rangle = 4 \pi G a^2 \langle \delta {T_\mathrm{m}}_0^0 \rangle\,.  \label{e:zero}
\ee
Here  $\langle \cdot \rangle$ is a spatial average taken with the unperturbed volume element.
We assume that the only inhomogeneous source of stress-energy is  nonrelativistic matter. Employing a particle description we can define a
``bare'' comoving number-density perturbation $\delta$ as
$
1 + \delta = \frac{dN}{d^3\vx}/ \left\langle \frac{dN}{d^3\vx} \right\rangle\, .
$
With this definition, the physical, or ``dressed'' energy density perturbation, within our approximation can be written as
\be
\label{e:dT}
 \delta {T_\mathrm{m}}_0^0 = \rho_0 \left[1 - \left(1 + 3 \Phi + \frac{1}{2} \overline{v^2}\right) \left(1 + \delta\right)\right] \, ,
\ee
where $\rho_0$ is the background matter density, and $\overline{v^2}$ denotes a phase space integral over the local velocity distribution. This
approximation takes into account the first corrections coming from the kinetic energy and the perturbation of the volume, including the
homogeneous perturbation $\Phi_0$. Here $\Phi_0$ can also be understood as a perturbative correction to the scale factor $a$, from the
averaged stress-energy of the perturbations which is ignored at the background level. It therefore induces a correction to the
expansion rate, $\Delta \HH$, which in our approximation is given by $\Delta \HH = -\Phi_0'$. The quantitative estimation of this correction and
its dependence on the amount of small scale inhomogeneities present in the simulation is the main aim of this paper. The Hubble rate we consider here
is that associated with the rest-frame of the gravitational field which has 4-velocity $n^\mu$~-- in this frame the magnetic Weyl curvature consists
only of induced vector and tensor modes, i.e.\ purely non-Newtonian terms~\cite{2011JCAP...03..029U}. By contrast, the Hubble rate associated with
`averaged observers' corresponding to a macroscopic fluid element has a 4-velocity tilted with respect to $n^\mu$, and corrections could be two or three orders
of magnitude larger than the results we find here~\cite{2011JCAP...03..029U}. Note that eq.~(\ref{order-4}) is actually related to the latter definition.
In the nonlinear regime of structure formation the relation between the two gauges becomes highly non-trivial. Even though we think this is an interesting issue,
this is not what we focus on in this letter.

Inserting Eq.~(\ref{e:dT}) in Eq.~(\ref{e:zero}), and using $\Omega_m(z)=8\pi G\rho_0(a)/(3H^2(a))$, one finds
\begin{multline}
3\HH\Phi'_0 + \frac{9}{2} \HH^2 \Omega_m(z) \Phi_0 = \frac{5}{2} \langle\Phi \nabla^2 \Phi\rangle \\- \frac{3}{2} \HH^2 \Omega_m(z) \left[3 \langle \Phi \delta \rangle + \frac{1}{2} \left\langle \left(1 + \delta\right) \overline{v^2}\right\rangle\right]\,.
\end{multline}

A simple interpretation emerges if we replace the quantities on the right-hand side by their Newtonian counterparts, i.e.\
$\Phi \rightarrow \psi_N$, $\delta \rightarrow \delta_N$. 
If we define the Newtonian total kinetic energy and total potential (binding) energy, respectively, as
$
2T = \sum_{i=0}^{N} {m_i} v_i^2 \, ,\quad
2U = \sum_{i=0}^{N} {m_i} \psi_N(x_i) \, ,
$
we obtain
\be\label{e:phi0}
2 \Phi'_0 + 3 \HH \Omega_m(z) \Phi_0 = -\HH \Omega_m(z) \frac{T + U}{M} \, ,
\ee
where $M = \sum_{i=0}^N m_i$ is the total rest mass, and we used Poisson's equation to relate $\delta_N$ and $\psi_N$. The perturbation of
the expansion rate is therefore driven by the  mean kinetic and binding energy densities of the matter particles, which both are ignored at the
background level.

In Newtonian cosmology, $T$ and $U$ obey the Layzer-Irvine equation~\cite{Irvine:1961,1963ApJ...138..174L},
$
\label{e:LE}
T' + U' + \HH \left(2 T + U\right) = 0.
$
This implies that as soon as most of the matter has accumulated in virialized structures, such that the virial
relation $2 T = -U$ holds to a good approximation, the total energy $T + U$ is conserved and $\Phi_0$ approaches a constant,
$\Phi_0 \ra -(T + U)/(3M)$. The correction to the expansion rate, $\Delta \HH = -\Phi_0'$ therefore approaches zero in the virial limit. Any corrections to this are a consequence
of relativistic effects.

We have solved Eq.~(\ref{e:phi0}) numerically using for the right hand side the results from different 3D simulations carried out with
\textsc{Gadget-2}~\cite{Springel:2005nw,Springel:2005mi,Springel:2000yr}. Our relativistic 1D simulations, on the other hand, directly solve for the second-order
potential $\Phi$ as described in~\cite{Adamek:2013wja}, and we can obtain $\Phi_0$ directly. Even though, not surprisingly,
the amplitudes are different, qualitatively the 1D and 3D results agree.

\begin{figure}[tb]
\includegraphics[width=0.95\columnwidth]{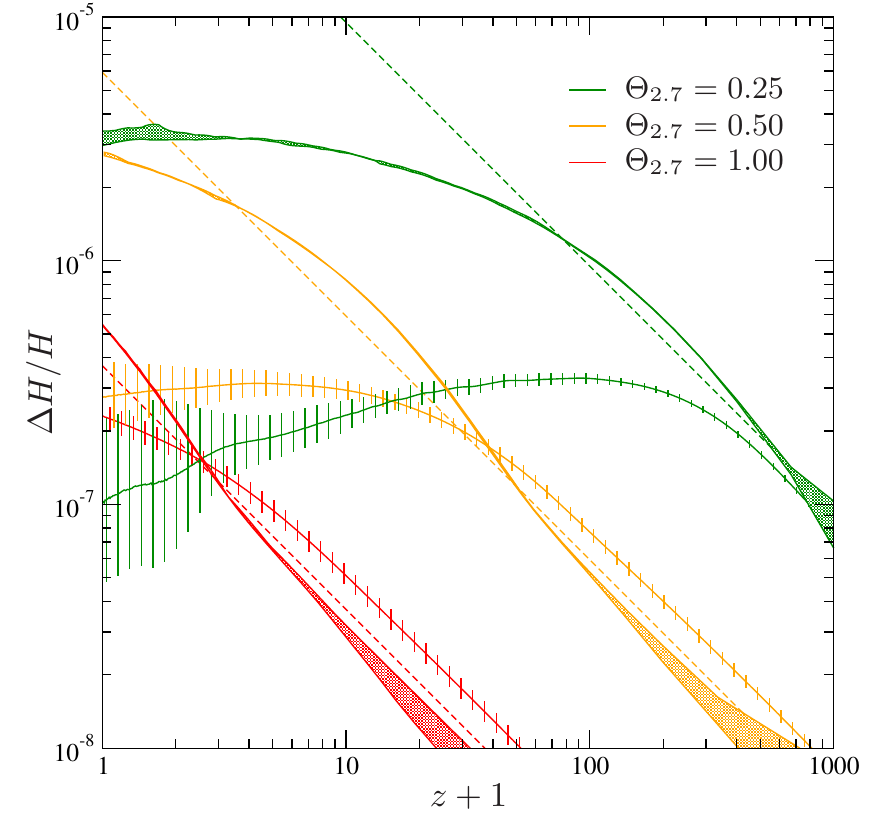}
\caption{\label{fig:br2} The perturbation of the Hubble rate from backreaction for different values of $\Theta_{2.7}$ in Einstein--de~Sitter universes. 
The solid lines with error bars show the ensemble average
and realization scatter for plane-symmetric relativistic simulations, whereas the shaded areas show the post-Newtonian estimate obtained from three 3D Newtonian N-body
simulations. The latter is based on the estimate on the average kinetic and potential energy of the particles, the size of the shaded regions giving a rough indication for the uncertainty which
is mainly caused by numerical deviations from the energy constraint, i.e.\ the Layzer-Irvine equation. Dashed lines are the  prediction from second
order perturbation theory. For $\Theta_{2.7}=0.25$ the backreaction $\De H/H$ stabilizes
roughly around $z_{\rm st}\simeq 55$ while for $\Theta_{2.7}=0.5$ this happens at $z_{\rm st}\simeq 2.5$. Note that for
 $\Theta_{2.7}=1$ the time of stabilization is actually in the future, $z_{\rm st}<0$.
In the 1D simulations the stabilization occurs earlier and at a lower amplitude.
}
\end{figure}

In Fig.~\ref{fig:br2} we plot the perturbation of the Hubble parameter as a function of redshift for different values of $\Theta_{2.7}$ which
is related to the equality scale by $k_{\rm eq}(\Theta_{2.7}) = \Theta_{2.7}^{-2}k_{\rm eq}(1)$. In order not to mistake the effects from
non-linearities by those of a cosmological constant, which leads to a decay of the gravitational potential due to the more rapid expansion, we
have simulated pure flat matter models (Einstein--de Sitter). For $\Theta_{2.7}=1$ we have $k_{\rm eq}(1)=\HH_{\rm eq}(1)=2H_0\Om_r^{-1/2}= 0.1h^2/$Mpc and 
$1+z_{\rm eq}(\Theta_{2.7}) = (1+z_{\rm eq}(1)) \Theta_{2.7}^{-4}$. Assuming that clustering leads to the stabilization of $\De H/H=-\Phi_0'/\mathcal{H}$, we expect
that the redshift when this happens is proportional to $z_{\rm eq}$
and therefore scales as $ \Theta_{2.7}^{-4}$. This is reasonably well verified in Fig.~\ref{fig:br2}.

\begin{figure}[tb]
\includegraphics[width=0.95\columnwidth]{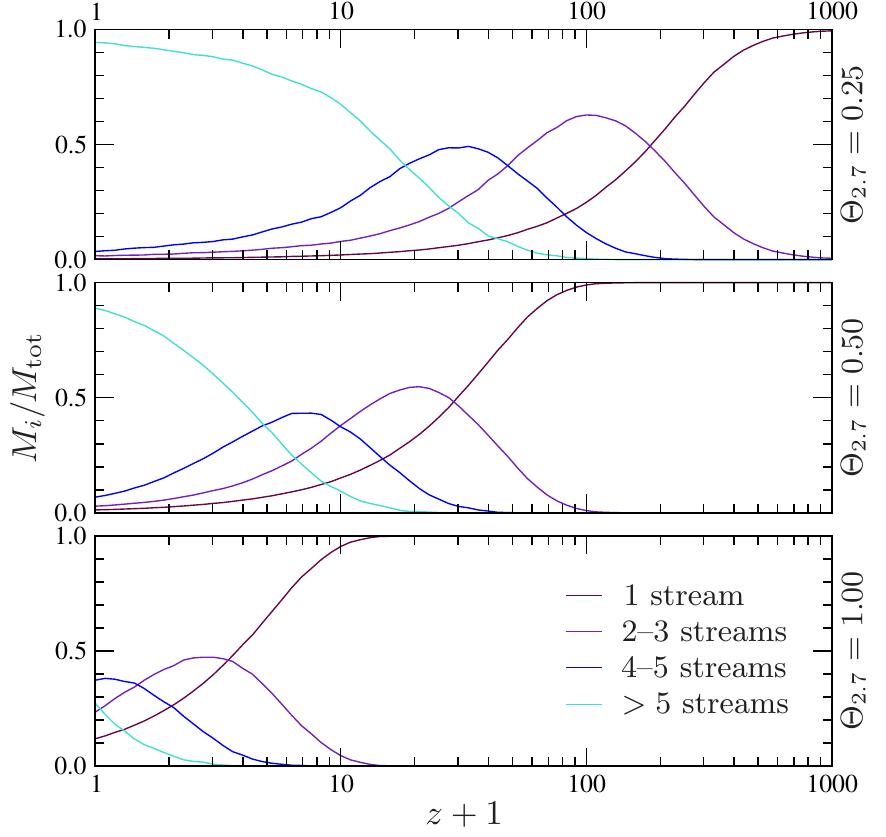}
\caption{\label{fig:mass} As structure formation proceeds, more and more mass is accumulated in non-linear structures. The figure shows the evolution with time of the fraction of mass which
resides in regions which contain $n$ velocity streams for the plane symmetric simulations. For $n>1$ these regions have undergone shell crossing and this fraction is zero initially,
since at the beginning the matter perturbations are fully in the linear regime. Perturbation theory is expected to be a poor description when a significant proportion ($\sim 50\%$, say) of the
matter is in regions which contain more than one stream.}
\end{figure}

As further indication for the progress of structure formation we plot in Fig.~\ref{fig:mass} the mass fraction of the particles which are contained in regions where $n$ velocity streams overlap.
For $n>1$ this means that shell crossing  has occurred and structure formation has entered the non-linear regime.  Fig.~\ref{fig:mass} shows that this happens around $z\sim 200$ for
$\Theta_{2.7}=0.25$ while shell crossing only becomes relevant at $z\sim 3$ for the simulations with $\Theta_{2.7}=1$.

We also studied the impact of the UV cutoff, which is implemented in the simulations because of their finite resolution. The amplitude of the backreaction effect $\Delta H / H$
increases slightly with better mass resolution, but the dependence on the cutoff is very mild. 

The results shown in Fig.~\ref{fig:br2} for the 3D case are from three simulations with $512^3$ particles. In the plane symmetric case we are able to vary the cutoff in a much larger range, and the results shown are fully converged. In this case, the large scatter between realizations is caused by the finite volume. Fluctuations are enhanced by the fact that,
as opposed to 3D, there exists only a single mode for each given $k$. It should be noted that the realization scatter (i.e.\ cosmic variance) is not insignificant also in 3D. In particular, we find
that it is larger than other effects, e.g.\ the influence of mass resolution, gravitational softening length and other simulation parameters.

Our interpretation of these findings is that once `stable clustering' is established and most structures have formed and decoupled from the Hubble
flow, the Hubble flow just proceeds (nearly) as before and the structures on small scales are irrelevant. 
On larger scales, structure formation is still ongoing and the virial limit is only reached asymptotically.

\paragraph{Discussion and conclusions}

We have shown that, contrary to the expectations from perturbation theory, clustering does not induce large 
changes in the expansion rate. The contribution to backreaction from a given scale decays once the scale has entered the regime of stable clustering,
i.e.\ once the non-linear structures have virialized. In the real Universe, this stable clustering progresses to larger and larger scales as
time goes on until the Universe becomes $\La$-dominated, after which linear perturbations no longer grow and no further scales
enter the non-linear regime.

This result indicates that backreaction never becomes large, as the formation of non-linear structures does not accelerate the deviation from
the averaged behavior on large scales. Instead backreaction appears to be reduced with the onset of non-linear structure
formation. If this behavior of the perturbed Hubble rate is representative,
relativistic backreaction effects, while certainly being present and non-negligible for precision cosmology
with future large surveys, cannot explain the observed accelerated expansion of the Universe. 

Although our results and arguments are suggesting strongly that backreaction does not significantly affect the background, they are not yet fully conclusive. Two areas
especially need improvement. Firstly, we have not yet  run a fully relativistic 3D simulation. Instead we used a relativistic
plane-symmetric simulation and in addition computed the metric and relativistic effects based on the particle phase-space distribution from a
standard 3D Newtonian N-body simulation. Although the results from the two approaches agree qualitatively, it would be desirable to repeat the
analysis with a relativistic 3D simulation. We are planning to accomplish this task in the future. Secondly, it would be preferable to consider directly observables like distances to quantify the impact of backreaction.

\paragraph{Acknowledgments}
 JA, RD and MK acknowledge financial support from the Swiss NSF. CC is funded by the National Research Foundation (South Africa). Part of the numerical calculations for
this work were performed on the Andromeda cluster of the University of Geneva.

\bibliographystyle{utcaps}
\bibliography{julian}

\end{document}